\documentclass[%
 preprint,
showpacs,preprintnumbers,
 amsmath,amssymb,
aps,
]{revtex4-1}

\usepackage{graphicx}
\usepackage{dcolumn}
\usepackage{bm}

\begin{document}

\preprint{}

\title{Magnetic and Structural Properties of the Iron Oxychalcogenides La$_{2}$O$_{2}$Fe$_{2}$O\emph{M}$_{2}$ (\emph{M}= S, Se)}

\author{B. Freelon,$^{1*}$  Z. Yamani,$^{2}$ Ian Swainson,$^{3}$ R. Flacau,$^{2}$ Yu Hao Liu,$^{4}$ L. Craco,$^{5}$ M. S. Laad,$^{6}$ Meng Wang,$^{7}$ 
Jiaqi Chen,$^{8}$ R. J. Birgeneau,$^{9,10,11}$ and Minghu Fang$^{8}$}
\email[]{byron.freelon@louisville.edu}
\affiliation{$^1$ Department of Physics, University of Louisville, Louisville, KY 40208, USA \\
$^2$Canadian Nuclear Laboratories, Chalk River Laboratories, Chalk River, ON, K0J 1J0, Canada\\
$^3$Physics Section, International Atomic Energy Agency, Vienna International Centre, PO Box 100, 1400 Vienna, Austria\\
$^4$Department of Materials Science and Engineering, University of Illinois Urbana Champaign, Urbana, Illinois, 61801\\
$^5$Instituto de Fisica, Universidade Federal de Mato Grosso, 78060-900, Cuiaba, MT, Brazil \\
$^6$The Institute of Mathematical Sciences, C.I.T. Campus, 
Chennai 600 113, India \\
$^7$School of Physics, Sun Yat-Sen University, Guangzhou, 510275, P. R. China\\
$^8$Department of Physics, Zhejiang University, Hangzhou 310027, P. R. China\\
$^9$Department of Physics, University of California, Berkeley, California 94720, USA\\
$^{10}$Materials Science Division, Lawrence Berkeley National Laboratory, Berkeley, California 94720, USA\\
$^{11}$Department of Materials Science and Engineering, University of California, Berkeley, California 94720, USA
}




\date{\today}

\begin{abstract}
We present the results of structural and magnetic phase comparisons of  the iron oxychalcogenides La$_{2}$O$_{2}$Fe$_{2}$O\emph{M}$_{2}$ (\emph{M} = S, Se). Elastic neutron scattering reveals that \emph{M} = S and Se have similar nuclear structures at room and low temperatures.  We find that both
 materials obtain antiferromagnetic ordering at a Neel temperature \begin{math}T_{N}\end{math} 90.1 $\pm$ 0.16 K and 107.2 $\pm$ 0.06 K for \emph{M}= Se and S, respectively.  The magnetic arrangements of \emph{M} = S, Se are obtained through Rietveld refinement.  We find
the order parameter exponent $\beta$ to be 0.129 $\pm$ 0.006 for \emph{M} = Se and 0.133 $\pm$ 0.007 for \emph{M} = S.  Each of these values is near the Ising symmetry value of 1/8.  This suggests that although lattice and electronic structural modifications result from chalcogen exchange, the nature of the magnetic interactions is similar in these materials. 
\begin{description}
\item[PACS numbers]  71.27.+a, 25.40.Dn, 78.70.Nx, 

\end{description}
\end{abstract}

\maketitle


\section{\label{sec:level1}INTRODUCTION}
The discovery of superconductivity in iron pnictide (Fe$Pn$) compounds has generated considerable interest because these materials seem to be the only current alternative to the cuprates for comparably high transition temperatures to the superconducting state.  As with the cuprates, superconductivity in the iron pnictides appears by doping electrons or holes into a magnetically ordered parent compound.  An important question is whether the iron pnictide parent compounds are on the verge of a metal insulator transition.~\cite{0953-8984-23-22-223201,1742-6596-449-1-012025}  In order to establish the so-called incipient Mott scenario~\cite{Dai4118} in iron pnictides, it is important to identify the Mott insulating portion of phase diagram of these materials.   One way to drive an Fe pnictide into the Mott insulating phase is by reducing the electron kinetic energy $t$ and increasing the electron correlation interaction $U$. 

The iron oxychalcogenides La$_{2}$O$_{2}$Fe$_{2}$O$M$$_{2}$  \emph{M}  = (S, Se) provide a case study  in this approach.  These particular iron oxychalcogenides are parent compounds that have a composition such that the nominal valence of Fe is 2+ and contain an Fe square lattice, which is similar to, but expanded relative to the iron pnictides. La$_{2}$O$_{2}$Fe$_{2}$O(S, Se)$_{2}$ were first reported to have insulating properties by Mayer.~\cite{ANIEBACK:ANIE199216451}  In addition, the crystal structure contains tetragonally ordered, Fe$Pn$-like Fe planes for which chalcogens alternate above and below the iron atoms. Oxygen atoms are  contained in the \emph{RE} layers remniscent of the charge resevoirs of high-\begin{math}T_{c}\end{math} cuprates.\cite{PhysRevLett.104.216405,ANIEBACK:ANIE199216451}  Specific attention has
been given to iron oxychalcogenides ~\cite{{0953-8984-28-45-453001},{PhysRevB.94.075150}, {0953-8984-26-14-145602},{PhysRevB.95.174441}} by investigators seeking to discover new iron-based materials in which high-\begin{math}T_{c}\end{math} superconductivity might be obtained by doping the Mott insulating state.~\cite{PhysRevLett.104.216405,PhysRevB.86.195133,LangJSNM2015}   





In addition, the electronic behavior of La$_{2}$O$_{2}$Fe$_{2}$O(S, Se)$_{2}$ has been investigated.  Zhu \emph{et. al.}, using local density  to dynamical mean-field theory (LDA + DMFT), predicted~\cite{PhysRevLett.104.216405} Mott insulating behavior and band narrowing of La$_{2}$O$_{2}$Fe$_{2}$O(S, Se)$_{2}$.  Bulk transport and magnetic measurements, in addition to resonant inelastic x-ray scattering (RIXS) and soft x-ray absorption spectroscopy~\cite{PhysRevB.92.155139}, provided experimental evidence that the systems are indeed Mott insulators.~\cite{PhysRevLett.104.216405}    Band narrowing within La$_{2}$O$_{2}$Fe$_{2}$O(S, Se)$_{2}$ has been proposed to lead to a Mott insulating state as well as enhanced electron correlation effects.  While exhibiting strongly correlated Mott insulating behavior, La$_{2}$O$_{2}$Fe$_{2}$O(S, Se)$_{2}$ may offer tunability of their electronic properties near a metal-insulator transition (MIT).  In Mott insulators, it has been observed that sizable electronic correlations drives new physical effects upon doping (electron or hole) and other external perturbations.~\cite{ImadaReview}  They can induce a range of interesting behavior including a pseudogap regime as in the case of Na$_{2}$Fe$_{2}$OSe$_{2}$~\cite{0953-8984-26-14-145602} ~\cite{Phillips2005}  or orbital-selective incoherent states that naturally yield co-existent insulating and bad metallic states as in cuprates or iron-pnictides.~\cite{Boebinger2011,SembaPRL2001,SimonsonPRB2011}  Furthermore, iron oxychalcogenides can be tuned by transition-metal or
chalcogen substitution~\cite{PhysRevLett.104.216405}, to produce novel electronic and magnetic phases at low temperature.
The substitution of S and Se has been shown to
alter the character of electronic partial density of states in this material class.~\cite{PhysRevLett.104.216405,PhysRevB.92.155139, 0953-8984-28-14-145701}  Even the presence of superconductivity in the iron oxychalcogenides has been of interest such that the effects of F-doping in ${\mathrm{La}}_{2}{\mathrm{O}}_{3-x}{\mathrm{F}}_{x}{\mathrm{Fe}}_{2}{\mathrm{Se}}_{2}$ and the substitution of Mn for Fe in ${\mathrm{La}}_{2}{\mathrm{O}}_{2}{\mathrm{Fe}}_{1-x}{\mathrm{Mn}}_{x}{\mathrm{Se}}_{2}$  have been investigated; however, no HTSC was observed. ~\cite{LangJSNM2015}~\cite{PhysRevB.86.125122}~\cite{Landsgesell2014232}
In addition to interesting electronic properties, studies of the magnetic behavior of iron oxychalcogenides have been pursued. \emph{A}$_{2}$F$_{2}$TM$_{2}$O(\emph{M})$_{2}$ where \emph{A} = (Ba, Sr) have been the subject of recent studies ~\cite{doi:10.1021/cm1035453, doi:10.1021/ja711139g,doi:10.1021/ja109007g,PhysRevLett.104.216405, PhysRevB.86.125122, 0953-8984-25-8-086004, Liu2015263, Landsgesell2014232} which showed that these materials order antiferromagnetically.  Further, Ba$_{2}$F$_{2}$Fe$_{2}$OSe$_{2}$ was proposed to be an example of a compound with a frustrated AFM checkerboard spin lattice.~\cite{doi:10.1021/ja711139g}  Other oxychalcogenides exhibit the onset of AFM ordering 2D short-range magnetic correlation well above \emph{T$_{N}$}.~\cite{PhysRevB.84.205212}  Stock \emph{et. al.} recently reviewed magnetic frustration and spin fluctuations in La$_{2}$O$_{2}$Fe$_{2}$OSe$_{2}$~\cite{0953-8984-28-45-453001} and other oxychalcogenides.  

In this work, we  study and compare the structural and
magnetic properties of the iron oxychalcogenides  La$_{2}$O$_{2}$Fe$_{2}$OS$_{2}$ and La$_{2}$O$_{2}$Fe$_{2}$OSe$_{2}$ using neutron powder
diffraction (NPD). Our focus is on powder materials since single-crystalline samples remain difficult to produce.  We measure the nuclear and magnetic Bragg scattering
intensity as a function of temperature and we examine the structural distinctions between the two parent compounds at room and low temperatures.  Section II provides the experimental details of the neutron and transport measurements.  Section III gives the results of structural and magnetic diffraction of \emph{M}= (S, Se).  In addition, we discuss the magnetic structure and the magnetic order parameter behavior which reveal Ising symmetry in both materials.  We provide a discussion of our findings within the context of specific magnetic exchange interactions and relative to other oxychalcogenides reports in the literature.  Finally, our results are discussed in light of some theoretical findings that have been reported on La$_{2}$O$_{2}$Fe$_{2}$O(S, Se)$_{2}$ systems. 


\section{\label{sec:level1}EXPeriment} 

The samples studied here ${\mathrm{La}}_{2}{\mathrm{O}}_{2}{\mathrm{Fe}}_{2}{\mathrm{O}}{\mathrm{(S,Se)}}_{2}$ have nominal compositions and were prepared by a conventional solid-state reaction method using high purity ${\mathrm{La}}_{2}{\mathrm{O}}_{3}$, S, Se and Fe powders as starting materials. The powders were mixed in the stoichiometric ratios and carefully ground. Subsequently, the powders were pressed into pellets and then heated in an evacuated quartz tube at $1030^\circ\mathrm{C}$ for 3 days; this process was repeated three times.  The samples were confirmed to be of a single phase by the laboratory X-ray powder diffraction measurements. ~\cite{ANIEBACK:ANIE199216451} 


Neutron powder diffraction (NPD) experiments were performed using the C2 high-resolution diffractometer at the  NRU reactor at Chalk River Canadian Nuclear Labs.  Room temperature measurements were conducted with approximately 3 g of finely ground powder of both ${\mathrm{La}}_{2}{\mathrm{O}}_{2}{\mathrm{Fe}}_{2}{\mathrm{S}}_{2}$ and ${\mathrm{La}}_{2}{\mathrm{O}}_{2}{\mathrm{Fe}}_{2}{\mathrm{Se}}_{2}$.  The samples were contained in vanadium cannisters sealed with indium gaskets under an atmosphere of He exchange gas. The low temperature NPD measurements were conducted using the same cannisters. All handling of the powders was performed inside a He glovebox.   The C2 diffractometer is equipped with an 800 wire position sensitive detector covering a range of 80 degrees. Data were collected in the angular range from $5^\circ\mathrm{}$ to $117^\circ\mathrm{}$ 2$\theta$ using a Si (5 3 1) monochromator at wavelengths $\lambda$ of 1.33 $\AA$ and 2.37 $\AA$.  Because ${\lambda}$ is similar in scale as the atomic spacing, the incident neutrons can be Bragg diffracted by nuclear positions.  Neutrons have zero-charge and a fermionic $S$ = 1/2, the resulting magnetic dipole moment of the neutron interacts with unpaired electrons to reveal magnetic ordering in solid materials.   
Rietveld analysis of the nuclear diffraction data estimated the samples to contain less than 1.2\% and 1.3\% of impurity phases in ${\mathrm{La}}_{2}{\mathrm{O}}_{2}{\mathrm{Fe}}_{2}{\mathrm{(S,Se)}}_{2}$, respectively.   

\begin{figure}
\includegraphics[width=4in]{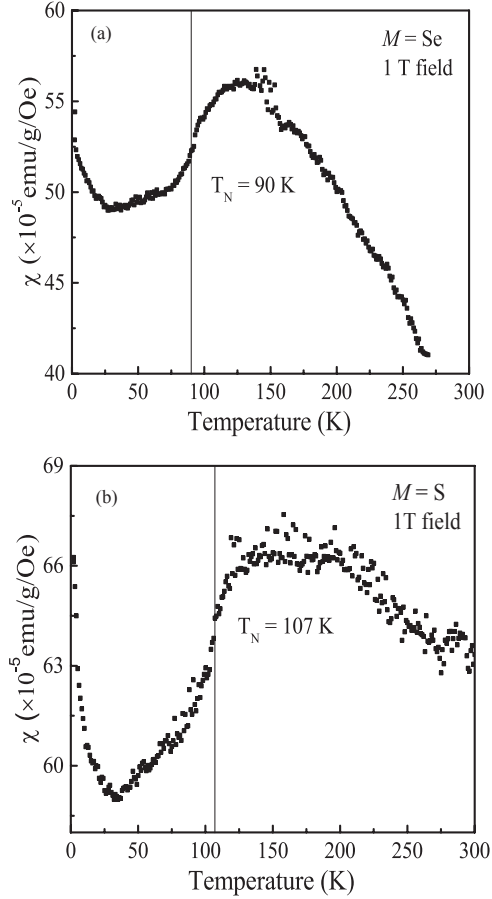}
\caption{(Color online) (a) The magnetic susceptibility of La$_{2}$OFe$_{2}$O$_{2}$\emph{M}$_{2}$ a) \emph{M}=Se and b)  \emph{M}=S.  
}
\label{fig:susceptibility}
\end{figure}

Resistivity versus temperature data for \emph{M} = S and Se have been published in Refs. ~\cite{PhysRevLett.104.216405} and ~\cite{PhysRevB.92.155139}.  \begin{math}\bf{M}\end{math} is defined as the magnetization per unit volume and \begin{math}\bf{H}\end{math} is the applied magnetic field (1 T in our measurement setup).  The magnetic susceptibility $d$\begin{math}\bf{M}\end{math}/$d$\begin{math}\bf{H}\end{math} \, as a function of temperature is shown in Fig. ~\ref{fig:susceptibility}. The data were collected on powder samples with $H$ = 1 T during warming using a Magnetic Properties Measurement System (MPMS) manufactured by Quantum Design, Inc.  The susceptibility data are similar to what is expected from 2D AFM samples except that there are Curie tails at low temperatures.  This could indicate the presence of a small concentration of paramagnetic impurities.  The Curie tails were not fitted since they do not affect the susceptibility curves in the Neel regions.

\section{\label{sec:level1}RESULTS}
\subsection{Nuclear Structure: \emph{M}= S, Se}

\begin{figure}
\includegraphics[width=4in]{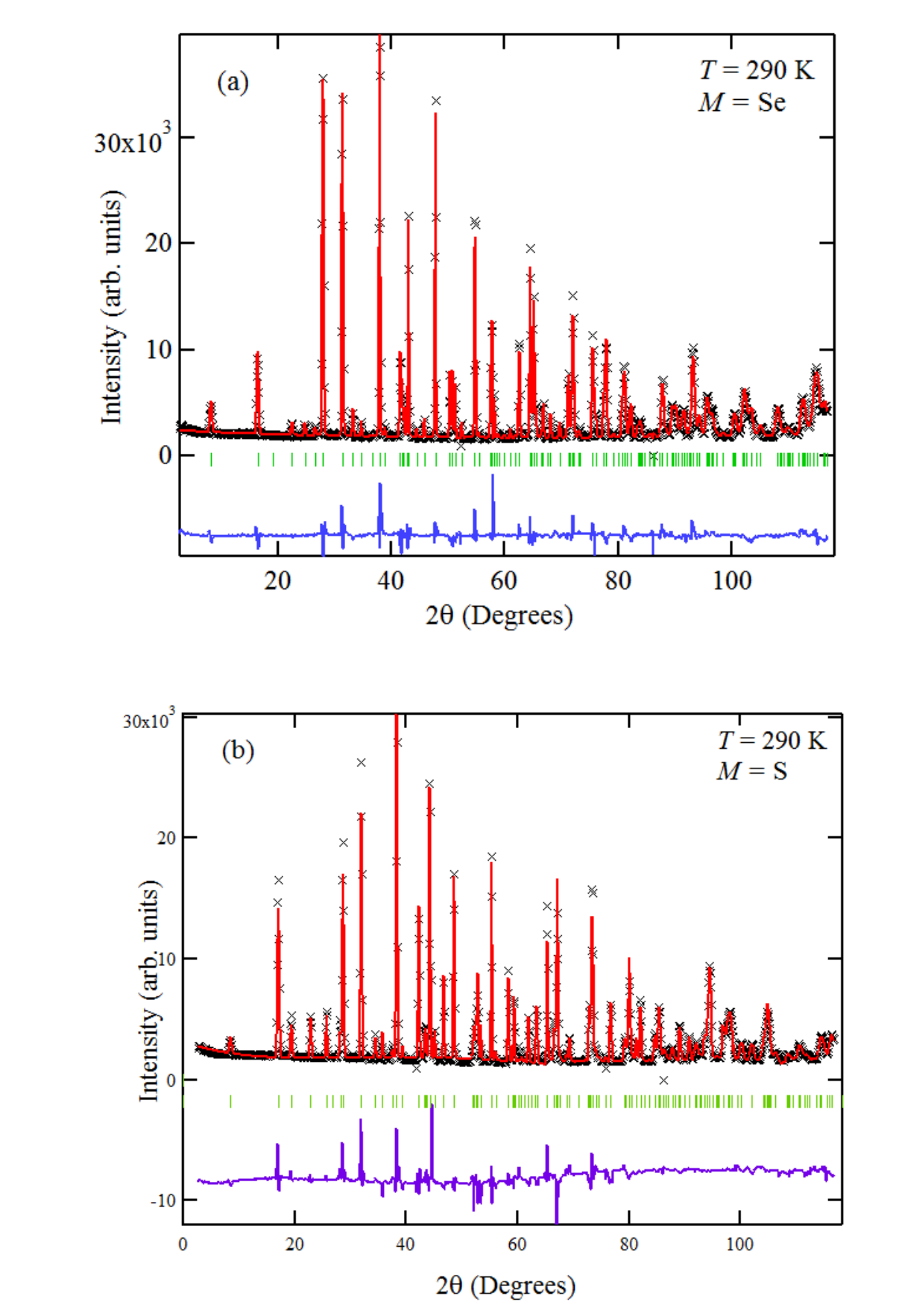}
\caption{(Color online) Rietveld refinement profiles using (a) \emph{M} = Se and (b) \emph{M} = S at 290 K,  data collected using the C2 diffractometer with wavelength $\lambda$ =1.33 $\AA$. The  data were refined using the space group $\emph{I}4/\emph{mmm}$.   Observed and calculated patterns are shown in red and black, respectively, with the difference profile (blue) and nuclear Bragg peak positions shown as blue given vertical tick marks.}
\label{fig:powdiffdata}
\end{figure}

Room temperature data was collected for crystal structure refinement in order to avoid magnetic Bragg peak contributions in analyzing the structural details of these materials. Fig. ~\ref{fig:powdiffdata} (a) and (b) shows the results of Rietveld structural refinement of (a) \emph{M}= S and (b) \emph{M}= Se at 290 K. The crystal structure refinement of  the  powder  neutron  diffraction  data was carried out using the FullProf~\cite{RODRIGUEZCARVAJAL199355} program suite, with the use of its internal  tables  for  scattering  lengths.     
These data are consistent with previous reports on \emph{M} = Se. ~\cite{PhysRevB.81.214433} 

\begin{figure}[!t]
\includegraphics[width=4.1in]{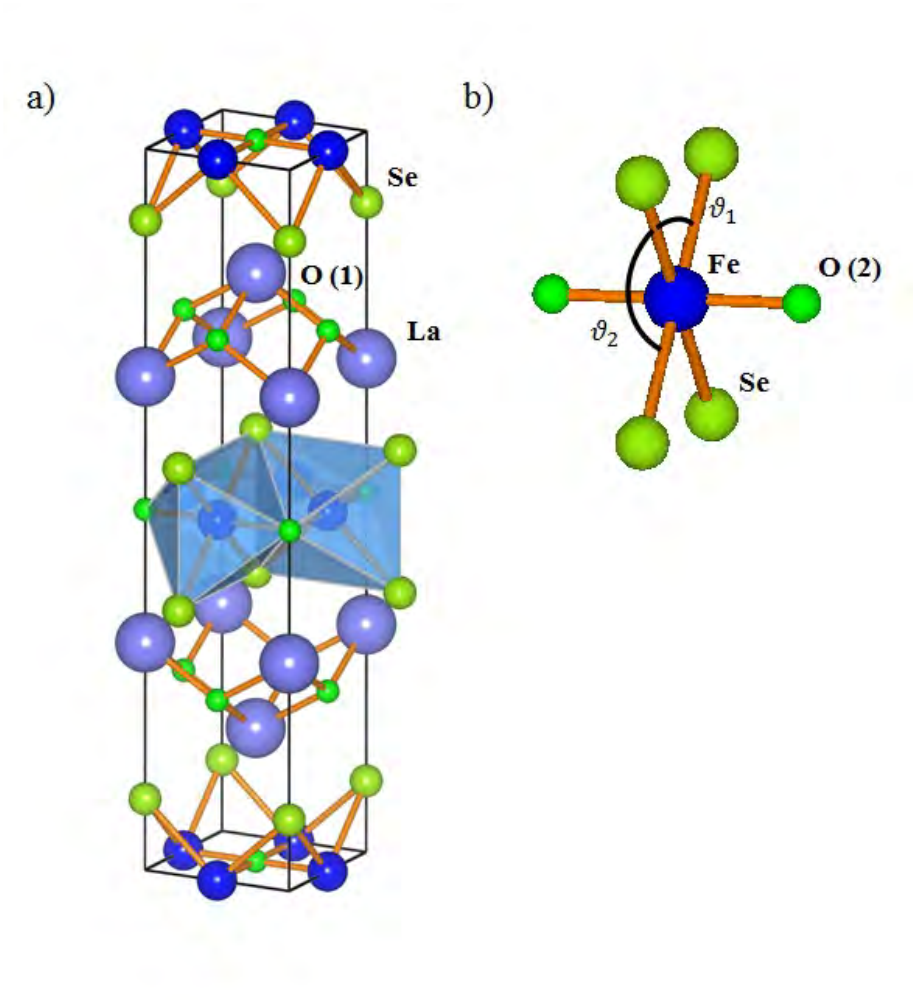}
\caption{(Color online) The crystal structure of La$_{2}$O$_{2}$Fe$_{2}$O(S,Se)$_{2}$ is shown in (a). The coordination geometry an Fe atom is shown in (b).  The angles ${\mathrm{\theta}}_{1}$ and ${\mathrm{\theta}}_{2}$ are described in the text.}
\label{atomic_structure}
\end{figure}

A direct comparison of the Rietveld refinement parameters of \emph{M}=  (S, Se) is given in Table I. 
\begin{table}
\begin{tabular}{ccc}\toprule
 & La$_{2}$OFe$_{2}$O$_{2}$Se$_{2}$ & La$_{2}$OFe$_{2}$O$_{2}$S$_{2}$\\
\hline
\multicolumn{2}{c}{} \\
\emph{a} [\AA]  	 					& 4.08778(5) 	& 	4.04539(5)\\
\emph{c} [\AA]      					&18.6005(3)	&	17.9036(3)      \\
 V[$\AA^3$]							&310.816(8)	&	292.997(8)       \\
4\emph{e} La(\emph{c})				&0.1843(1)		&	0.1811(1)      \\
4\emph{e} \emph{M}(\emph{c})       & 0.0964(1)	& 	0.0933(3)      \\
La B(\AA)     							&0.21(5) 		&	0.30(4)      \\
Fe B(\AA)    							&0.46(3)		&	0.42(4)      \\
O(1) B(\AA)    							&0.44(5)		&	0.40(4)      \\
O(2) B(\AA)    							&0.91(8)		&	0.89(7)      \\
\emph{M} B(\AA)								&0.38(4)		&	0.26(8)	\\
\hline
\end{tabular}
\caption{\label{tab:example}Refined parameters for La$_{2}$OFe$_{2}$O$_{2}$Se$_{2}$  La$_{2}$OFe$_{2}$O$_{2}$S$_{2}$ at  290 K.}
\end{table}
Fe unit cell volume of \emph{M}= Se is larger than that of \emph{M}= S.   This is reasonable given that the ionic radius of sulfur (100 pm) is smaller than selenium (115 pm) and we find that the  
 \begin{table*}[ht]
\footnotesize\baselineskip=8pt
\begin{center}
\begin{tabular}{lccc|cccccccc}
\hline 
\hline
 \multicolumn{2}{c}{La$_{2}$O$_{2}$Fe$_{2}$OSe$_{2}$}	& 	&&La$_{2}$O$_{2}$Fe$_{2}$OS$_{2}$\\
\hline
 Length & [\AA]$^{\mbox{b}}$  & 
  Bond Angle   &   [$^{\circ}$]  &  Length    &  [\AA]    &  Bond Angle    &  [$^{\circ}$]\\
\hline
 \emph{d}$_{Fe-Fe}$                             &  2.89050(3) & 
  Fe-O-Fe (2)  &  90.00  &  \emph{d}$_{Fe-Fe}$	&  2.85813(0)&  Fe-O-Fe (2) &  90.00\\
\emph{d}$_{Fe-O2}$                              &   2.04389(3) & 
  Fe-Se-Fe   &  64.086 &  \emph{d}$_{Fe-O2}$  &  2.02100(0)   &  Fe-S-Fe  &  64.919\\
\emph{d}$_{Fe-Se}$                               &  2.72400(4) &  
  Se-Fe-Se  &  97.237  &  \emph{d}$_{Fe-S}$ & 2.66264(0)  &  S-Fe-S &  98.755\\
\emph{d}$_{La-O1}$$^{\mbox{e}}$                &  2.38013(3)  & 
 La-O-La (1) & 118.348 &  \emph{d}$_{La-O1}$ &   2.33787(0)    &  La-O-La (1)   &  119.642\\
\emph{d}$_{La-Se}$$^{\mbox{e}}$                &  3.31832(4)  & 
 La-O-La (2) & --- &  \emph{d}$_{La-S}$ &   3.26261(0)    &  La-O-La (2)   &  ---\\
\hline
\end{tabular}
\end{center}
\caption{\label{tab:example}Interatomic distances and angles of  La$_{2}$O$_{2}$Fe$_{2}$O(S, Se)$_{2}$  at 290 K.}
\end{table*}
lattice parameters of La$_{2}$O$_{2}$Fe$_{2}$OS$_{2}$ (\emph{a} =4.04539$\AA$, \emph{c} = 17.9036$\AA$) are smaller than those of La$_{2}$O$_{2}$Fe$_{2}$OSe$_{2}$ (\emph{a} =4.0877$\AA$, \emph{c} = 18.6005$\AA$).  Compared to the Fe-Fe atomic distances $\it{d}$$_{Fe-Fe}$ values reported in LaFeOAs, the interatomic distances we obtained (in Table II) are larger by 1.2 \% and nearly 1.1\% for \emph{M}= Se and S, respectively.  Bond angles and atomic distances, extracted from Rietveld refinement parameters, are tabulated for comparison of \emph{M}= S and Se (see Table II).  Rietveld analysis yielded measurements for the bond angles subtended by Se-Fe-Se and S-Fe-S defined as ${\mathrm{\theta}}_{1}$ and ${\mathrm{\theta}}_{2}$, respectively.  These bonds define the  distortion of the Fe(S, Se)\begin{math}_{4}\end{math} squares contained in the Fe$_{2}$OS$_{4}$ and Fe$_{2}$OSe$_{4}$ octahedra ($\it{c.f.}$ Figure 3(a)), respectively.  For completeness Table III presents the atomic position of constituents of La$_{2}$O$_{2}$Fe$_{2}$O(S, Se)$_{2}$.

The atomic structure (see Fig.~\ref{atomic_structure}~(a)) of La$_{2}$O$_{2}$Fe$_{2}$O(S, Se)$_{2}$ crystallizes in the space group \emph{I}4/\emph{mmm} (No. 139).~\cite{ANIEBACK:ANIE199216451}  La$_{2}$O$_{2}$Fe$_{2}$O(S, Se)$_{2}$ contain anti-CuO$_{2}$-type square, planar stacks such that [La$_{2}$O$_{2}$]$^{2+}$ layers and [Fe$_{2}$O]$^{2+}$ layers are separated by (S, Se)$^{2-}$ anions.      The Fe$^{2+}$ cations are linked through two in-plane Oxygen O(2) anions as well as four out-of-plane (S$^{2-}$, Se$^{2-}$) anions.  The Fe atoms are tetrahedrally coordinated with (S, Se) atoms alternatingly located, above or below, the center of the Fe-O plaquettes; therefore, the Fe-\emph{M} layers are not flat.  These \emph{D}$_{2\it{h}}$ point symmetry octahedra are face sharing ($\it{c.f.}$ Fig.~\ref{atomic_structure}~(a)) such that the shared face is intersected by the Fe-Fe nearest-neighbor line-of-sight.~\cite{doi:10.1021/cm1035453} 

Here we list the angles ${\mathrm{\theta}}_{1}$ and ${\mathrm{\theta}}_{2}$ (see Fig.~\ref{atomic_structure}~(b)) for \emph{M} = S and Se along with that of other oxychalcogenides.  Specifically, La$_{2}$O$_{2}$Fe$_{2}$OS$_{2}$, $98.7^\circ\mathrm{}$ and $81.3^\circ\mathrm{}$; La$_{2}$O$_{2}$Fe$_{2}$OSe$_{2}$, $97.2^\circ\mathrm{}$ and $82.8^\circ\mathrm{}$; Nd$_{2}$O$_{2}$Fe$_{2}$OSe$_{2}$, $96.1^\circ\mathrm{}$ and $83.9^\circ\mathrm{}$ ~\cite{0953-8984-22-34-346003}; La$_{2}$O$_{2}$Co$_{2}$OSe$_{2}$, $98.7^\circ\mathrm{}$ and $81.3^\circ\mathrm{}$ ~\cite{Fuwa20101698}; Sr$_{2}$Ti$_{2}$F$_{2}$OAs$_{2}$, $96.3^\circ\mathrm{}$ and $83.7^\circ\mathrm{}$; Sr$_{2}$Ti$_{2}$F$_{2}$OSb$_{2}$, $91.0^\circ\mathrm{}$ and $89.0^\circ\mathrm{}$ ~\cite{Charkin20117344};  Sr$_{2}$Fe$_{2}$F$_{2}$OS$_{2}$, $100.2^\circ\mathrm{}$ and $79.8^\circ\mathrm{}$; Ba$_{2}$Fe$_{2}$F$_{2}$OS$_{2}$, $102.2^\circ\mathrm{}$ and $77.8^\circ\mathrm{}$; Sr$_{2}$Fe$_{2}$F$_{2}$OSe$_{2}$, $97.2^\circ\mathrm{}$ and $82.8^\circ\mathrm{}$.~\cite{doi:10.1021/ja711139g}  A comparison of the \begin{math}\theta_{1}\end{math} values for \textit{M} = S, Se indicate that the S atoms are closer to the iron plane than Se chalcogens.  This Fe-S distance results in greater octahedral distortion for the \textit{M} = S material as compared to \textit{M} = Se.~\cite{Liu2015272}

\begin{table*}
\footnotesize\baselineskip=8pt
\begin{center}
\begin{tabular}{lccccccccccc}
\hline 
\hline
La$_{2}$O$_{2}$Fe$_{2}$OSe$_{2}$	&&&&& La$_{2}$O$_{2}$Fe$_{2}$OS$_{2}$\\
\hline
   Atom & Site  & 
  x   &   y  &  z  &  Atom    &  Site    &  x    &  y     &  z &  \\
\hline
\textbf{La}                               & 4e & 
  0.5000  &  0.5000 &  0.1844 & \textbf{La}  & 4e &  0.5000 &  0.5000 & 0.18105 \\
\textbf{Fe}                       & 4c & 
  0.5000 &  0.0000 &  0.0000 &   \textbf{Fe} & 4c &   0.5000 &  0.0000   &  0.0000 \\
\textbf{Se}                       & 4e & 
  0.0000 &  0.0000 &  0.0968 & \textbf{S} & 4c & 0.0000 &  0.0000   &  0.0933 \\
\textbf{O1}                       & 4d & 
  0.5000 &  0.0000 &  0.2500 &   \textbf{O1} & 4d &   0.5000 &  0.0000   &  0.2500 \\
\textbf{O2}                      & 2b & 
  0.5000 &  0.5000 &  0.0000 &   \textbf{O2} & 2b &   0.5000 &  0.5000   &  0.0000 \\
\hline
\end{tabular}
\end{center}
\caption{\label{tab:atompositions}Atomic positions of La$_{2}$O$_{2}$Fe$_{2}$O(S, Se)$_{2}$  at 290 K extracted from refined parameters.}
\end{table*}

Our low-temperature,  high-resolution, \emph{M} = (S, Se) powder diffraction data does not contain pattern changes or structural Bragg peak splittings that would indicate the occurrence of a thermally driven structural phase transition.  We do not observe the emergence of an atomic arrangement with lower symmetry as a function of temperature.  This is consistent with results for \emph{M} = Se reported by Free \emph{et. al.}~\cite{PhysRevB.81.214433}  Those authors noticed subtle temperature-dependent lattice behavior and atomic displacement parameter \begin{math}U_{33}\end{math} trends in La$_{2}$O$_{2}$Fe$_{2}$OSe$_{2}$.~\cite{PhysRevB.81.214433}  It was suggested that these features were, at best, weak indications a lowering of lattice symmetry.  It has been proposed that the absence of a structural phase transition is due to the reduction of magnetostructural coupling in La$_{2}$O$_{2}$Fe$_{2}$O(S,Se)$_{2}$ caused by structural disordering.~\cite{PhysRevB.81.214433}     


A more detailed study of the local (short-range) structure of La$_{2}$O$_{2}$Fe$_{2}$O(S,Se)$_{2}$ was performed in order to determine whether there is any localized structural arrangement as a function of temperature; a pulsed neutron scattering study of the local structure of La$_{2}$O$_{2}$Fe$_{2}$OSe$_{2}$ was conducted by extracting atomic position deviations from radial distribution function data.~\cite{kazumasa_horigane_local_2014} No local structure change from the low-temperature  \emph{I}/4\emph{mmm} symmetry was observed in these experiments.  Both Fuwa~\cite{0953-8984-22-34-346003} and Free ~\cite{PhysRevB.81.214433} suggested that the absence of structural phase transitions might be due to the lack of magnetostriction or magnetoelastic coupling in La$_{2}$O$_{2}$Fe$_{2}$OSe$_{2}$.~\cite{0953-8984-26-14-145602,PhysRevB.80.224506}

\subsection{Magnetic Structure: \emph{M}= S, Se}
\begin{figure}
\includegraphics[width=4in]{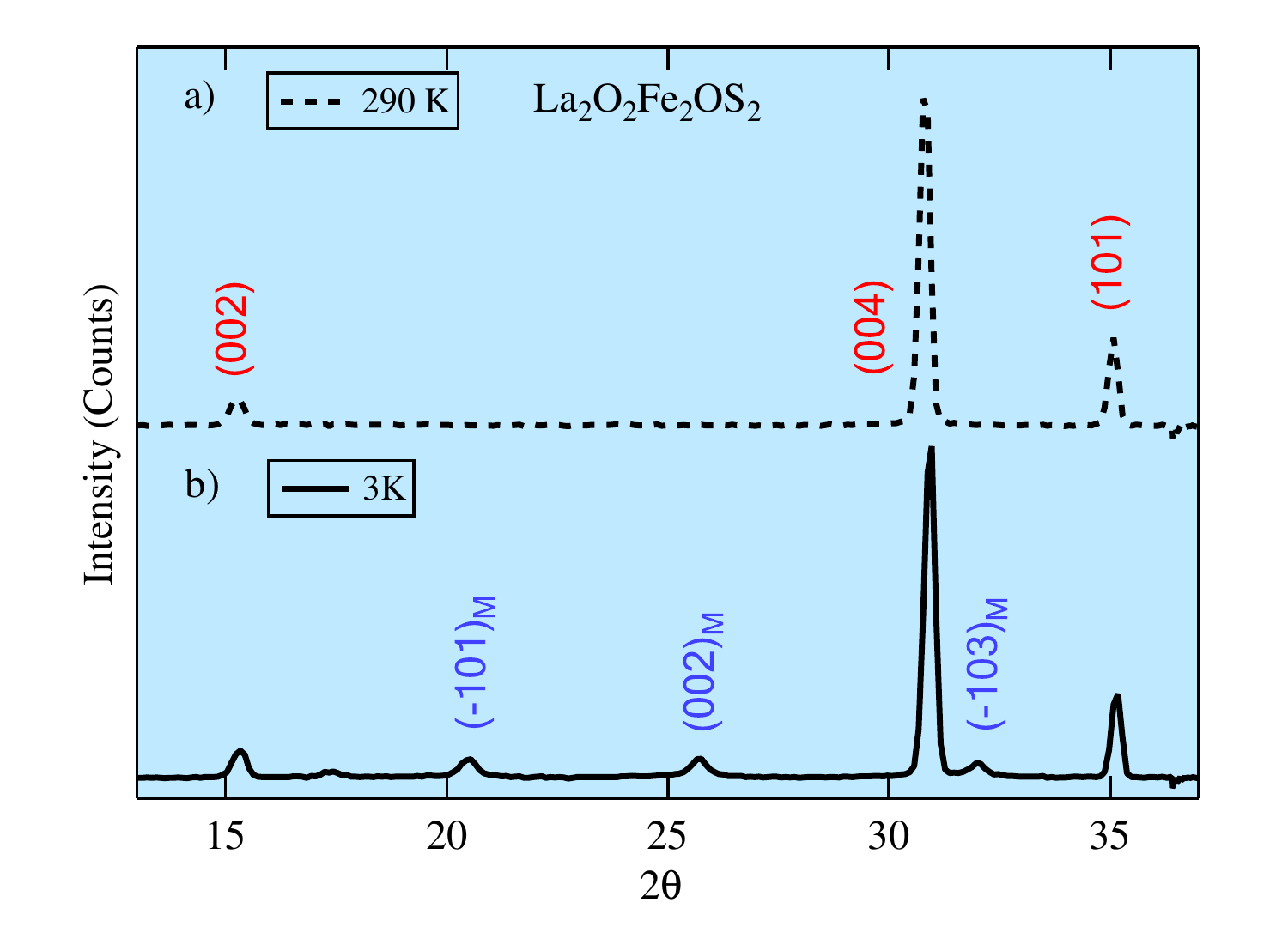}
\caption{(Color online) The neutron powder diffraction data for La$_{2}$OFe$_{2}$O$_{2}$S$_{2}$ is shown for a) 3 K and b) 290 K. Red (purple) labels indicate the ($HKL$) indices of the structural (magnetic) Bragg diffraction peaks.   
}
\label{fig:lowangledffxn}
\end{figure}

Upon cooling, extraneous intensity appeared in the diffraction profiles of \emph{M} = S and Se which we attributed to phase transition behavior.   These peaks were assigned a magnetic origin on the basis of their temperature dependence and the complete Rietveld refinement of the diffraction patterns. The Se end member has been well-characterized by Free and Evans~\cite{PhysRevB.81.214433} and we initially followed their approach by verifying that the AFM3 model provided the best fit to the magnetic structure of La$_{2}$O$_{2}$Fe$_{2}$OSe$_{2}$.  We used the SARAh suite of programs ~\cite{Wills2000680} to analyze the representations and provide the basis vectors for refinement with FullProf.~\cite{RODRIGUEZCARVAJAL199355,roisnel2001WinPLOTR} In close similarity to what was described previously for \emph{M} = Se, the magnetic cell of \emph{M} = S is commensurate and is doubled in \emph{a} and \emph{c} with respect to the structural cell. The magnetic ordering in La$_{2}$O$_{2}$Fe$_{2}$OS$_{2}$ is associated with an ordering vector \begin{math}\mathbf{k}=(1/2,0,1/2)\end{math}, and the single Fe site on $\lbrace1/2,0,0\rbrace$ in the nuclear \emph{I}4/$mmm$ cell is described by two distinct orbits governing the two $\lbrace1/2,0,0\rbrace$ and $\lbrace0,1/2,0\rbrace$ Fe sites that are independent in the magnetically ordered state.   In Fig.~\ref{fig:lowangledffxn} the low angle region  of the powder diffractograms of La$_{2}$O$_{2}$Fe$_{2}$OS$_{2}$ is shown at 3 K and 290 K.  The data are presented such that magnetic Bragg peaks \begin{math}\textbf{Q}_{M} = (-101), (002)\end{math} and \begin{math}(-103)\end{math} are seen (blue label) to develop at low-temperatures while the structural Bragg peaks are present (black) for both high- and low-temperatures.  

There are two independent Fe sites for both La$_{2}$O$_{2}$Fe$_{2}$O(S, Se)$_{2}$.  By performing a full Rietveld refinement and analysis of the neutron powder diffraction data, the ordered Fe$^{2+}$ moment of La$_{2}$O$_{2}$Fe$_{2}$OS$_{2}$ was determined to be 2.32(4) $\mu$$_{B}$ at 4 K; for La$_{2}$O$_{2}$Fe$_{2}$OSe$_{2}$ a range of values from 2.8 to 3.50(5) $\mu$$_{B}$ has been reported.~\cite{PhysRevB.89.100402}  We then found that the AFM3 model provided good fits for the structure of \emph{M} = S.  Based upon the refinement, the symmetry and orientation of the magnetic moments can be assigned to a \emph{I}4/\emph{mmm} symmetry for \emph{M} = S and Se. 
Thus the magnetic ground state is composed of the Fe ions in a high spin, non-collinear antiferromagnetically ordered state.    

\begin{figure}
\includegraphics[width=4in]{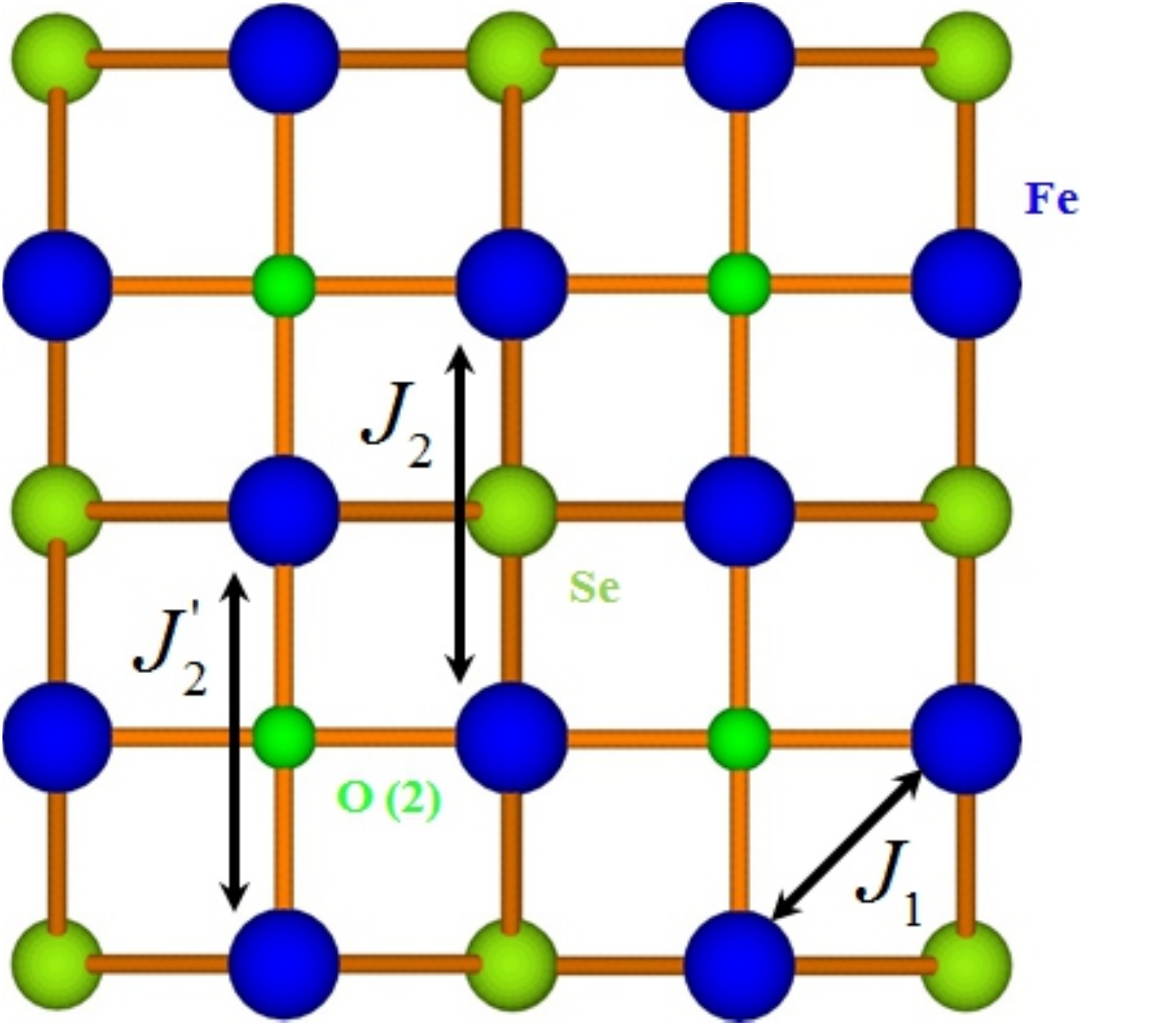}
\caption{(Color online)  Spin interactions $J_{1}$, $J_{2}$ and $J'_{2}$ used in modeling the magnetic behavior of La$_{2}$O$_{2}$Fe$_{2}$O(S,Se)$_{2}$.  These are the principle spin interactions contained in the Fe$_{2}$O(S, Se)$_{2}$ layer.  The details regarding each exchange constant are given within the text.}
\label{fig:magneticinteractions}
\end{figure}

In order to discuss the magnetic structure results obtained from La$_{2}$O$_{2}$Fe$_{2}$O(S,Se)$_{2}$, we summarize the most prominent description of the spin interactions in \emph{RE}-O$_{2}$Fe$_{2}$O\emph{M}$_{2}$ materials.  Several spin interaction labeling conventions can be found in the literature on La$_{2}$O$_{2}$Fe$_{2}$OSe$_{2}$; table IV lists them. We adopt the convention used in Refs. ~\cite{PhysRevLett.104.216405}, ~\cite{PhysRevB.81.214433} and ~\cite{PhysRevB.89.100402}.  The spin Hamiltonian for La$_{2}$O$_{2}$Fe$_{2}$O(S,Se)$_{2}$  has been modeled by  Zhu \emph{et. al} ~\cite{PhysRevLett.104.216405} using three interactions $J_{1}$, $J_{2}$ and $J'_{2}$.  $J_{1}$ has several contributions: (a) a face-sharing $64^\circ$ interaction between Fe-Se-Fe (b) an Fe-O-Fe $90^\circ$ interaction and (c) possibly an iron nearest neighbor (NN) contribution. $J_{2}$ is a next nearest-neighbor (NNN) interaction that consists of a ~$98^\circ$ edge-sharing term involving of Fe-Se-Fe contributions from two buckled Se atoms (iii) $J'_{2}$ is a NN, $180^\circ$ Fe-O-Fe interaction between the corner-sharing octahedra. 
Fig.~\ref{fig:magneticinteractions} provides a schematic description of these interactions.

\begin{figure}[!t]
\includegraphics[width=5in]{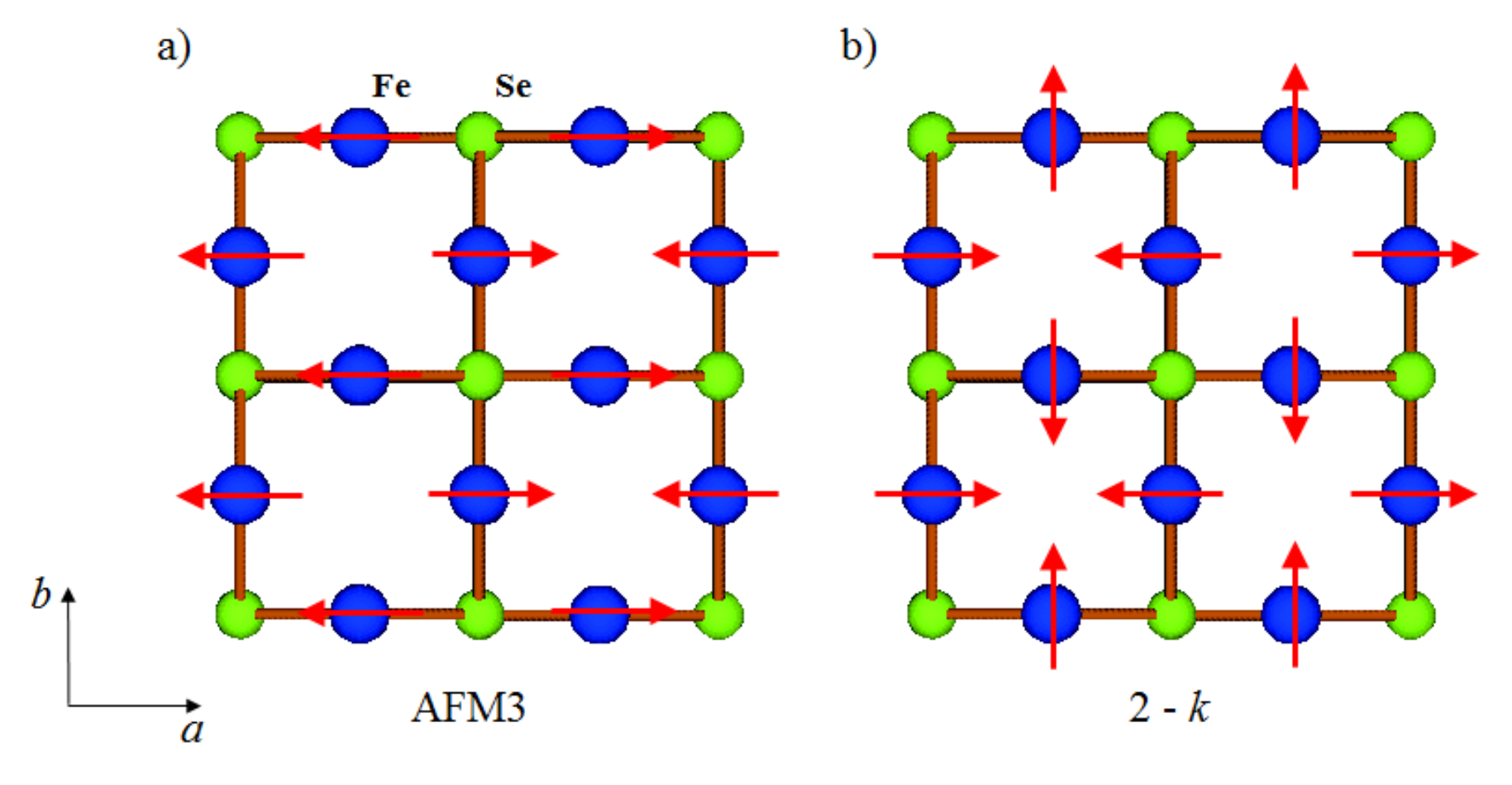}
\caption{(Color online)  The (a) collinear AFM3 and (b) non-collinear 2-\emph{k} magnetic structure models for La$_{2}$O$_{2}$Fe$_{2}$OSe$_{2}$.}
\label{fig:magneticstructuresAFM3_6}
\end{figure}

Using the interaction ($J_{1}, J_{2}, J'_{2}$) labeling, Zhu $\emph{et. al.}$ employed a generalized gradient approximation (GGA) + Coulomb energy \emph{U} calculation~\cite{PhysRevLett.104.216405} in order to determine that the magnetic ground state of La$_{2}$O$_{2}$Fe$_{2}$OSe$_{2}$ should obtain either the AFM1 or the AFM6 ($c.f.$ Fig. 6 in Ref.~\cite{PhysRevB.81.214433}) configuration depending on the value of $U$. By contrast, Free and Evans reported that elastic neutron scattering results indicated~\cite{PhysRevB.81.214433} that the magnetic moments of La$_{2}$O$_{2}$Fe$_{2}$OSe$_{2}$ should order in the AFM3 (see Fig.~\ref{fig:magneticstructuresAFM3_6}(a)) frustrated, collinear configuration similar to Fe$_{1.086}$Te. At variance with this finding, McCabe \emph{et. al.} concluded that inelastic neutron scattering (INS) results on La$_{2}$O$_{2}$Fe$_{2}$OSe$_{2}$ are consistent with a multi-component, non-collinear 2-\emph{k} magnetic structure shown in Fig.~\ref{fig:magneticstructuresAFM3_6}(b). The 2-\emph{k} structure is made up of 2 orthogonal stripes within the Fe$_{2}$O$M_{2}$ layers.  While the AFM3 configuration provided good fits to our \emph{M} = S and Se NPD data, as noted~\cite{PhysRevB.89.100402}, neutron powder diffraction can not distinguish between the various AFM models that have been proposed for La$_{2}$O$_{2}$Fe$_{2}$OSe$_{2}$.  Therefore the consistency between INS data and the 2-\emph{k} magnetic structure offer insight to understanding the magnetic structures of La$_{2}$O$_{2}$Fe$_{2}$O(S, Se)$_{2}$ given that the production of single-crystals is  difficult.  
In addition to neutron scattering experiments, nuclear magnetic resonance (NMR) measurements by Gunther \emph{et. al.} ~\cite{PhysRevB.90.184408-Marco-gunter} and Ref. ~\cite{arXivFreelon_SN} were interpreted to suggest that the 2-\emph{k} model is the appropriate description of the \emph{M} = Se.

Magnetic frustration has been addressed in La$_{2}$O$_{2}$Fe$_{2}$OSe$_{2}$ and other iron oxychalcogenides.  Some amount of frustration in La$_{2}$O$_{2}$Fe$_{2}$O(S, Se)$_{2}$ is to be expected given that there are FM ($J_{2}$) and AFM ($J'_{2}$) interactions associated with the FeO$M_{2}$ layer.  These competing interactions, in addition to magnetocrystalline anisotropy, create a frustrated spin environment in which the FeO$M_{2}$ layer has three principal competing magnetic interactions $J_{1}$, $J_{2}$ and $J'_{2}$.~\cite{Liu2015272} In  contrast to the AFM3 collinear frustrated model of Ref. ~\cite{PhysRevB.81.214433}, McCabe \emph{et. al.} suggested the 2-\emph{k} La$_{2}$O$_{2}$Fe$_{2}$OSe$_{2}$ structure to be weakly frustrated due to the magnetic configuration being collectively stabilized by the AFM $J'_{2}$ and FM $J_{2}$ interactions as well as the magnetocrystalline anisotropy.  Those authors  reported that both the magnetic frustration and the exchange coupling are weak in La$_{2}$O$_{2}$Fe$_{2}$OSe$_{2}$ compared to other iron oxychalcogenides.  The variance in these reports suggests that even though the extent of magnetic frustration in La$_{2}$O$_{2}$Fe$_{2}$OS$_{2}$ and La$_{2}$O$_{2}$Fe$_{2}$OSe$_{2}$ is not fully understood, however, it is still possible to compare the magnetic frustration of \emph{M} = S and Se.  As seen above, our structural refinements yield a smaller bond angle Fe-\emph{M}-Fe in \emph{M} = S than for \emph{M} = Se.  This is an indication that the \emph{M} = S \emph{NNN} distances are smaller and, therefore, have increased magnetic exchange interaction.  In addition, the \emph{NN} exchange of \emph{M} = S is increased relative to that of \emph{M} = Se.  It has been proposed~\cite{0953-8984-22-34-346003} that the \emph{NNN} AFM interaction $J'_{2}$ in Nd$_{2}$O$_{2}$Fe$_{2}$OSe$_{2}$ is due to the Goodenough-Kanamori rule.~\cite{doi:10.1143/PTP.17.177} This reasoning might explain the difference in Neel temperatures observed for \emph{M} = S and \emph{M} = Se.     
\begin{table}[h]
\caption{Magnetic Interaction Naming Conventions} 
\centering 
\begin{tabular}{c cccc}     
\hline\hline                    
Interaction&Description &Geomtery &Ref.~\cite{PhysRevLett.104.216405, PhysRevB.89.100402, PhysRevB.81.214433},[$^{\mbox{a}}$] &~\cite{PhysRevB.86.125122, doi:10.1021/ja711139g,Liu2015272}\\[0.5ex]   
\hline            
\emph{NNN}   & Fe-O-Fe & $180^\circ$ &  $J'_{2}$& $J_{1}$\\   
\emph{NN}     & TM-TM &in-plane &  $J_{1}$&$J_{3}$ \\[0.5ex] 
\emph{NNN} & TM-\emph{M}-TM &$\sim90^\circ$ &  $J_{2}$& $J_{2}$\\
\hline                          
\multicolumn{5}{l}{Notes:} The (next) nearest neighbor interactions are  (\emph{N})\emph{NN}\\
\multicolumn{5}{l}{$^{\mbox{a}}$ This work.} \\

\end{tabular}
\label{tab:hresult}
\end{table}

Several spin Hamiltonians have been introduced in order to address the magnetic behavior of materials thought to be within strong coupling limit \emph{e.g.} iron oxychalcogenides and iron alkaline selenides  $A$$_{0.8}$Fe$_{1.6}$Se$_{2}$ $\rightarrow$ $A$$_{2}$Fe$_{4}$Se$_{5}$ (referred to as "245s" where \emph{A} = Rb, Cs, K and Tl).~\cite{Si_Abrahams_NatRev2016,PhysRevB.92.165102,PhysRevB.95.205132,0953-8984-28-49-495702,PhysRevB.92.165102} Unlike the iron pnictides, the increased electron correlation of the  iron chalcogenides and iron alkaline selenides leads to narrower iron bandwidths.  Consequently, despite the absolute value of the Hund’s coupling being similar to that of the pnictides  ($J_{H}$ $\sim$ 0.7 eV),  its  role  is  more  pronounced, resulting in a larger spin $S$ = 2.    
Importantly,  INS experiments~\cite{PhysRevB.89.100402} have yielded $S$ = 2 for La$_{2}$O$_{2}$Fe$_{2}$OSe$_{2}$ and Rb$_{0.8}$Fe$_{1.5}$S$_{2}$~\cite{PhysRevB.92.079901}.  

Finally, the absence of La$_{2}$O$_{2}$Fe$_{2}$O(S,Se)$_{2}$ structural phase transitions can be compared to the iron pnictides, which actually undergo a structural phase transition from tetragonal to orthorhombic symmetry at a structural transition temperature ${T}_{S}$.  
In the case of Fe$Pn$, a structural phase transition is either concomitant with or immediately prior to an AFM phase transition.  The presence of ferro-orbital ordering of the $d_{xz,yz}$ states is intimately linked to the Fe$Pn$ structural phase transitions. Furthermore, iron-pnictide ferro-orbital ordering is associated with magnetic phase changes by virtue of spin-orbit coupling and Coulomb interaction.~\cite{PhysRevB.80.224506}  By contrast, the absence of ferro-orbital ordering in La$_{2}$O$_{2}$Fe$_{2}$O(S,Se)$_{2}$ may be due to the apparent non-degeneracy of the $d_{xz,yz}$ orbitals.~\cite{PhysRevB.92.155139} 
\subsection{Magnetic Order Parameter Critical Behavior}
\begin{figure}
\includegraphics[width=4in]{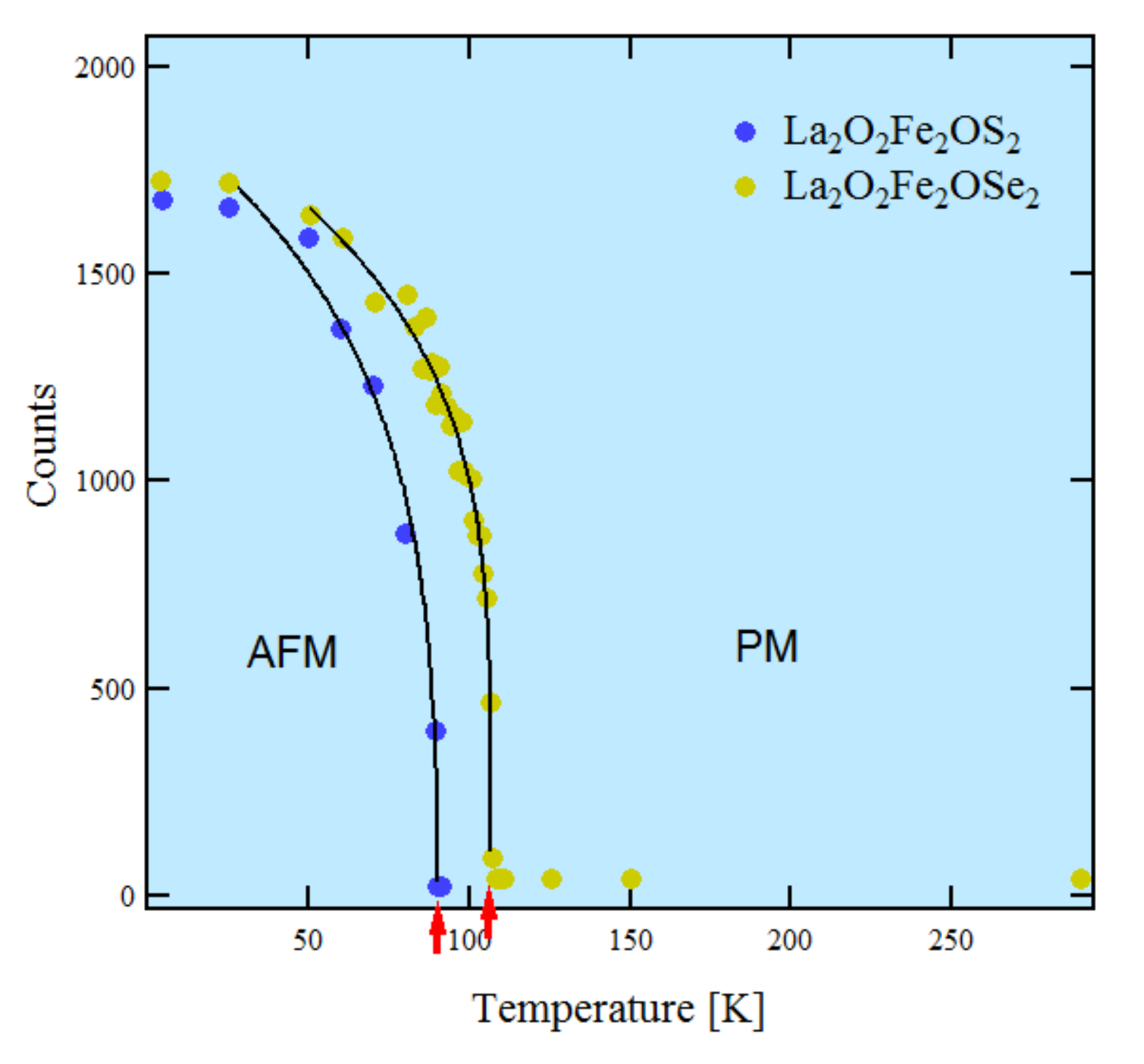}
\caption{(Color online)   The magnetic order parameter
is shown at the (1, 0, 3) magnetic Bragg reflection. The peak
intensity measured at \textbf{Q} = (1, 0, 3), plotted for both La$_{2}$O$_{2}$Fe$_{2}$O(S,Se)$_{2}$, is used as a measure of the magnetic order parameter $\phi^2$.}
\label{fig:MagOPfit}
\end{figure}

To determine the thermal dependence of the magnetic ordering behavior of \emph{M} = (S, Se), the intensity of the magnetic Bragg peak {\bf Q} = (-103) was measured over a temperature range of 300 K to 4 K in each material.  The peak intensity can be used as a measure of the magnetic order parameter squared  $\phi$$^{2}$.  The square of the magnetic order parameters of La$_{2}$O$_{2}$Fe$_{2}$O(S, Se)$_{2}$ are plotted in Fig.~\ref{fig:MagOPfit}.    La$_{2}$O$_{2}$Fe$_{2}$O(S,Se)$_{2}$ peak intensity data was fitted to the power-law functional form $\phi$$^{2}$\begin{math}(T/T_{N})= (1-T/T_{N})^{2\beta_{Fe}}\end{math}.~\cite{PhysRevB.79.184519}  $\beta_{Fe}$, the critical exponent, and \begin{math}T_N\end{math}, the N\`{e}el temperature, served as adjustable fit parameters.  Fits were applied over the temperature range 0.05 $\le$ \begin{math}T/T_{N}\le 1 \end{math} and yielded values for $\beta_{Fe}$ and \begin{math}T_{N}\end{math} of  0.129 $\pm$ 0.006 and 90.1(9) $\pm$ 0.16 K for \emph{M} = Se and 0.133 $\pm$ 0.007  and 107.2(6) $\pm$ 0.06 K in the case of \emph{M} = S.  These extracted values of $\beta_{Fe}$ are close to those reported for \emph{M} = Se in Ref.~\cite{PhysRevB.81.214433}.  Furthermore, the $\beta_{Fe}$ for both \emph{M} = S and Se are close to the Ising critical exponent $\beta_{Ising}$ value of 1/8.  This result indicates that the magnetic phase transitions in La$_{2}$O$_{2}$Fe$_{2}$OS$_{2}$ and La$_{2}$O$_{2}$Fe$_{2}$OSe$_{2}$ may be weakly first order in agreement with results obtained using M\"ossbauer spectroscopy.~\cite{arXivFreelon_SN} 
These critical exponent values also suggest that there are 2D Ising-like spin fluctuations near the critical point.  The similarity of the critical exponents is further indication that the magnetic phases of La$_{2}$O$_{2}$Fe$_{2}$OS$_{2}$ and La$_{2}$O$_{2}$Fe$_{2}$OSe$_{2}$ arise from similar magnetic interaction geometries.


\section{\label{sec:level1}CONCLUSIONS}
We have presented a comparison of the structural and magnetic properties of the homologues La$_{2}$O$_{2}$Fe$_{2}$O(S,Se)$_{2}$ based on bulk transport and neutron powder diffraction data.  Our motivation was to present a comparison of the structural and magnetic details of \emph{M}= S and Se as there had been no previously published, explicit comparison of these compounds.  Neutron powder diffraction indicates that the nuclear structures of La$_{2}$O$_{2}$Fe$_{2}$O(S,Se)$_{2}$ are similar to the structural character found in other oxychalcogenides with the main distinction being the difference in lattice size based on the atomic radii of the two chalcogens.  Nuclear Bragg diffraction data indicates that the FeO$_{2}$Se$_{4}$ and FeO$_{2}$S$_{4}$ octahedra have different sizes; this produces different magnitudes of distortion within the octahedra.  This distortion is expected to be related to the presence of the relatively high extent of electron correlation compared to the iron pnictides.  In addition, the distorted octahedra can diminish magnetoelastic coupling by precluding orbital ordering that is necessary to establish a link between the magnetic phase transition and a structural phase transformation.~\cite{Liu2015272} We did not observe structural phase transitions in the materials.  Nor did we see evidence of a nematic phase similar to that which exists in the iron pnictides.  However, observing only a magnetic phase transition from the high-temperature paramagnetic phase to a lower temperature AF phase, we used group theory and magnetic refinement methods to determine the magnetic structure of these materials. The magnetic structure of La$_{2}$O$_{2}$Fe$_{2}$O(S,Se)$_{2}$ was determined to be consistent with a non-collinear 2-\emph{k} configuration up to the basic limitations imposed by the use of powder samples.  2D Ising symmetry was determined for both La$_{2}$O$_{2}$Fe$_{2}$O(S,Se)$_{2}$.  We discussed models of frustrated magnetism and their relevance to metallic and insulating behavior iron oxychalcogenides. 

\section{\label{sec:level1}AKNOWLEDGEMENTS}
We are grateful to the technical staff at CNBC for excellent support. Work at Lawrence Berkeley National Laboratory and UC Berkeley was supported by the Office of Science,
Office of Basic Energy Sciences (BES), Materials Sciences and
Engineering Division of the U.S. Department of Energy (DOE) under
Contract No. DE-AC02-05-CH1231 within the Quantum Materials Program
(KC2202) and BES. The work at Zhejiang University was supported by the National 
Basic Research Program of China (973 Program) under Grants No. 2011CBA00103 
and 2012CB821404, the National Science Foundation of China (No. 11374261, 
11204059) and Zhejiang Provincial Natural Science Foundation of China 
(No. LQ12A04007). M.S.L thanks MPIPKS, Dresden for hospitality.  
L.C.'s work was supported by CNPq (Grant No. 307487/2014-8). L.C. also thanks
the Physical Chemistry Department at Technical University Dresden for 
hospitality.   

\bibliography{ChalkRiver_FeOCh1_References.bib}

\end{document}